\documentclass[sigplan,authorversion=true]{acmart}

\usepackage{booktabs} % For formal tables

\usepackage{balance}

\usepackage{graphicx}
\usepackage{import}

\begin{document}
\title[Blackchain: Scalable Accountable V2X Communication]{Blackchain: Scalability for Resource-Constrained Accountable Vehicle-to-X Communication}
\author{Rens W. van der Heijden}
\orcid{0000-0003-3280-1825}
\affiliation{%
  \institution{Institute of Distributed Systems\\ Ulm University}
  \city{Ulm} 
  \state{Germany} 
}
\email{rens.vanderheijden@uni-ulm.de}

\author{Felix Engelmann}
\orcid{0000-0001-9356-0231}
\affiliation{%
  \institution{Institute of Distributed Systems\\ Ulm University}
  \city{Ulm} 
  \state{Germany} 
}
\email{felix.engelmann@uni-ulm.de}

\author{David M\"odinger}
% \orcid{0000-0003-3280-1825}
\affiliation{%
  \institution{Institute of Distributed Systems\\ Ulm University}
  \city{Ulm} 
  \state{Germany} 
}
\email{david.moedinger@uni-ulm.de}

\author{Franziska Sch\"onig}
% \orcid{0000-0003-3280-1825}
\affiliation{%
	\institution{Ulm University}
	\city{Ulm} 
	\state{Germany} 
}
\email{franziska.schoenig@uni-ulm.de}

\author{Frank Kargl}
\orcid{0000-0003-3800-8369}
\affiliation{%
  \institution{Institute of Distributed Systems\\ Ulm University}
  \city{Ulm} 
  \state{Germany} 
}
\email{frank.kargl@uni-ulm.de}

\begin{abstract}
  In this paper, we propose a new Blockchain-based message and revocation accountability system called Blackchain.
  Combining a distributed ledger with existing mechanisms for security in V2X communication systems, we design a distributed event data recorder (EDR) that satisfies traditional accountability requirements by providing a compressed global state.
  Unlike previous approaches, our distributed ledger solution provides an accountable revocation mechanism without requiring trust in a single misbehavior authority, instead allowing a collaborative and transparent decision making process through Blackchain.
  This makes Blackchain an attractive alternative to existing solutions for revocation in a Security Credential Management System (SCMS), which suffer from the traditional disadvantages of PKIs, notably including centralized trust.
  Our proposal becomes scalable through the use of hierarchical consensus: individual vehicles dynamically create clusters, which then provide their consensus decisions as input for road-side units (RSUs), which in turn publish their results to misbehavior authorities.
  This authority, which is traditionally a single entity in the SCMS, responsible for the integrity of the entire V2X network, is now a set of authorities that transparently perform a revocation, whose result is then published in a global Blackchain state.
  This state can be used to prevent the issuance of certificates to previously malicious users, and also prevents the authority from misbehaving through the transparency implied by a global system state.
\end{abstract}

%
% The code below should be generated by the tool at
% http://dl.acm.org/ccs.cfm
% Please copy and paste the code instead of the example below. 
%
\begin{CCSXML}
<ccs2012>
<concept>
<concept_id>10003033.10003083.10003014</concept_id>
<concept_desc>Networks~Network security</concept_desc>
<concept_significance>500</concept_significance>
</concept>
<concept>
<concept_id>10002978.10003014</concept_id>
<concept_desc>Security and privacy~Network security</concept_desc>
<concept_significance>300</concept_significance>
</concept>
  <concept>
    <concept_id>10003033.10003039.10003051.10003052</concept_id>
    <concept_desc>Networks~Peer-to-peer protocols</concept_desc>
    <concept_significance>300</concept_significance>
  </concept>
</ccs2012>
\end{CCSXML}

\ccsdesc[500]{Networks~Network security}
\ccsdesc[300]{Security and privacy~Network security}
\ccsdesc[300]{Networks~Peer-to-peer protocols}

\keywords{Distributed Ledger, VANET, Accountability}

%\begin{teaserfigure}
%  \includegraphics[width=\textwidth]{sampleteaser}
%  \caption{This is a teaser}
%  \label{fig:teaser}
%\end{teaserfigure}

\copyrightyear{2017}
\acmYear{2017}
\setcopyright{acmlicensed}
\acmConference[SERIAL'17]{SERIAL'17: ScalablE and Resilient InfrAstructures for distributed Ledgers}{December 11--15, 2017}{Las Vegas, NV, USA}
\acmBooktitle{SERIAL'17: SERIAL'17: ScalablE and Resilient InfrAstructures for distributed Ledgers, December 11--15, 2017, Las Vegas, NV, USA}
\acmPrice{15.00}
\acmDOI{10.1145/3152824.3152828}
\acmISBN{978-1-4503-5173-7/17/12}

\maketitle

% The default list of authors is too long for headers}
\renewcommand{\shortauthors}{R. W. vd Heijden, F. Engelmann, D. M\"odinger, F. Sch\"onig, F. Kargl}

\section{Introduction}
\label{sec:intro}

An event data recorder (EDR), the car equivalent of a flight recorder, can be used for a multitude of applications, e.g., forensic accident reconstruction~\cite{Raya-Securing,Kopylova-Reconstruction} and misbehavior detection~\cite{vanderHeijden-Survey}.
Although EDRs are easy to implement for in-vehicle communication, where the amount of events is limited, the required append-only and write-once semantics (i.e., it should not be possible to change logs) makes it non-trivial to implement for vehicle-to-vehicle communication.
The most notable disadvantage is that more expensive tamper-resistant hardware is required to either store the EDR log directly, or implement and protect cryptographic means to protect a log stored in a file, an issue caused mainly by the much higher message volume in V2X communication.
In addition, de-duplication of the different EDR logs across vehicles is not possible, i.e., when reporting misbehavior, the complete logs of all vehicles (at the very least for the corresponding time frame where misbehavior occurred) are required.

To tackle these problems, this paper examines the use of distributed ledgers (DL) for this application.
Distributed ledgers are distributed data storages, which provide an append-only semantic to the participants.
%This allows us to employ known techniques of data de-duplication and tamper-proofing the data.
The most well known implementation of a distributed ledger, Bitcoin~\cite{nakamoto2009bitcoin}, provides the consensus, and therefore tamper-proofing, by restricting the rate of data write through a proof-of-work mechanism.
However, a naive adoption of this ledger is not suitable for our scenario, because it does not scale to the message frequency encountered by an in-vehicle EDR.
Multiple works on scaling the transaction throughput with the help of off-chain state channels~\cite{Poon-lightning} or sharding~\cite{Luu-Sharding} help to improve the performance in the current setting, but are not optimised for our setting, coping with delays and significant churn. 

Therefore we propose a permissioned Blockchain, with which we can build a hierarchical byzantine fault tolerant consensus.
On the lowest layer, cars form clusters and agree on a state change which is propagated to a road side unit (RSU).
As there are too many RSUs deployed to reach a global consensus efficiently, smaller RSU groups form and aggregate a partial state.
The fixed set of transaction issuers allow for a weighted consensus and efficient, distributed mining process.

In recent years, with concrete implementation plans for V2X communication systems, researchers have started to look more closely at the design of a real-world public key infrastructure (PKI) and the multitude of requirements in such a system.
Most recent designs, such as that proposed by Whyte~et~al.~\cite{Whyte-SCMS}, include a misbehavior authority, in addition to a standard certificate revocation component as in regular PKIs.
This authority is responsible for accepting misbehavior reports, processing them according to some fixed algorithm, and revoking any vehicles that show malicious behavior.
In the real world, it is likely that not just one, but several of such SCMSs will be deployed by competing entities (either vehicle manufacturers or countries~\cite{Raya-Securing}).

To make the revocation process more transparent, and to reduce the trust necessary in any one SCMS, we propose the use of a distributed ledger for accountability.
This system will use the transactions received from the RSUs, backed by the permissioned Blockchain described above, and publish them in a public, permissionless Blockchain, for public verifiability.
This provides accountability between MAs, as well as towards the users of the system (i.e., the vehicle owners), due to the publicly verifiable nature of this chain.
This process aims to provide global revocation of misbehaving vehicles; existing proposals all require centralized trust for this process.
Although there is a lot of work on how to locally revoke malicious vehicles~\cite{Bilogrevic-OREN,Lui-Limits}, global revocation without central authorities is a challenge that has received limited attention so far. 

In the remainder of this paper, we introduce the conceptual foundations of our proposal.
Specifically, we describe a detailed system model, including privacy and attacker models, in Section~\ref{sec:sysmodel}.
Section \ref{sec:blackchain} then describes our Blackchain proposal and some possible attacks on our base system.
We finally discuss the implications of these ideas for distributed ledgers and misbehavior detection research in Section~\ref{sec:conclusion}.

\section{System and Attacker Model}
\label{sec:sysmodel}
Vehicular ad-hoc networks (VANETs) consist of vehicles and road-side units (RSUs), equipped with wireless communication modules.
Unlike traditional wireless networks, VANETs are primarily based on broadcast communication: vehicles periodically broadcast beacons, containing application-relevant information such as position, speed, heading, and some meta-data.
Applications of VANETs vary from crash avoidance to finding fastest routes and fuel and road efficiency applications, which can potentially be combined with self-driving vehicles to further increase performance.

Communication typically uses the IEEE 802.11 standard, with a range between 300 and 1000 meters; many authors propose more advanced communication patterns on top of this.
RSUs are typically assumed to be available in some locations only (e.g., attached to traffic lights), but provide an intermittent link to the Internet for all vehicles.
Some research suggests that the current work in 5G cellular communication may provide more permanent Internet connectivity, although this may be costly for users; a heterogeneous network using both technologies is a current hot topic in this community~\cite{Tung-Heterogeneous}.
For this paper, we focus on the case of clustering, where vehicles communicate with others in communication range directly, but a cluster head (CH) is responsible for communication with other clusters.
For an overview of clustering techniques, we refer interested readers to a recent survey by Cooper~et~al.~\cite{Cooper-Clustering}.

In VANETs, security plays an important role, due to the lives dependent on the communication.
Unlike existing IT infrastructures, the main focus of security lies on integrity and availability, rather than confidentiality; it is generally assumed that the message contents are not encrypted, since any vehicle needs access to this content for any VANET application to provide any real benefit.
Message integrity is generally protected through signed messages, where each vehicle possesses a number of authentic public keys from a vehicular public key infrastructure (VPKI).
This PKI has one additional requirement: each vehicle should receive multiple certificates, in order to protect user privacy; this is important, because each broadcast message contains a signature signed with a public key, and the corresponding certificate is attached to the message.
The certificates are referred to as pseudonyms, which contain a cryptographically protected link to the users' long-term certificate.
To prevent attackers from using all certificates at the same time, these pseudonyms have limited validity periods that only overlap partially; in addition, there is a revocation protocol that reveals the cryptographic link if misbehavior is detected.

\begin{figure*}
	\centering
	\includegraphics[width=0.9\columnwidth]{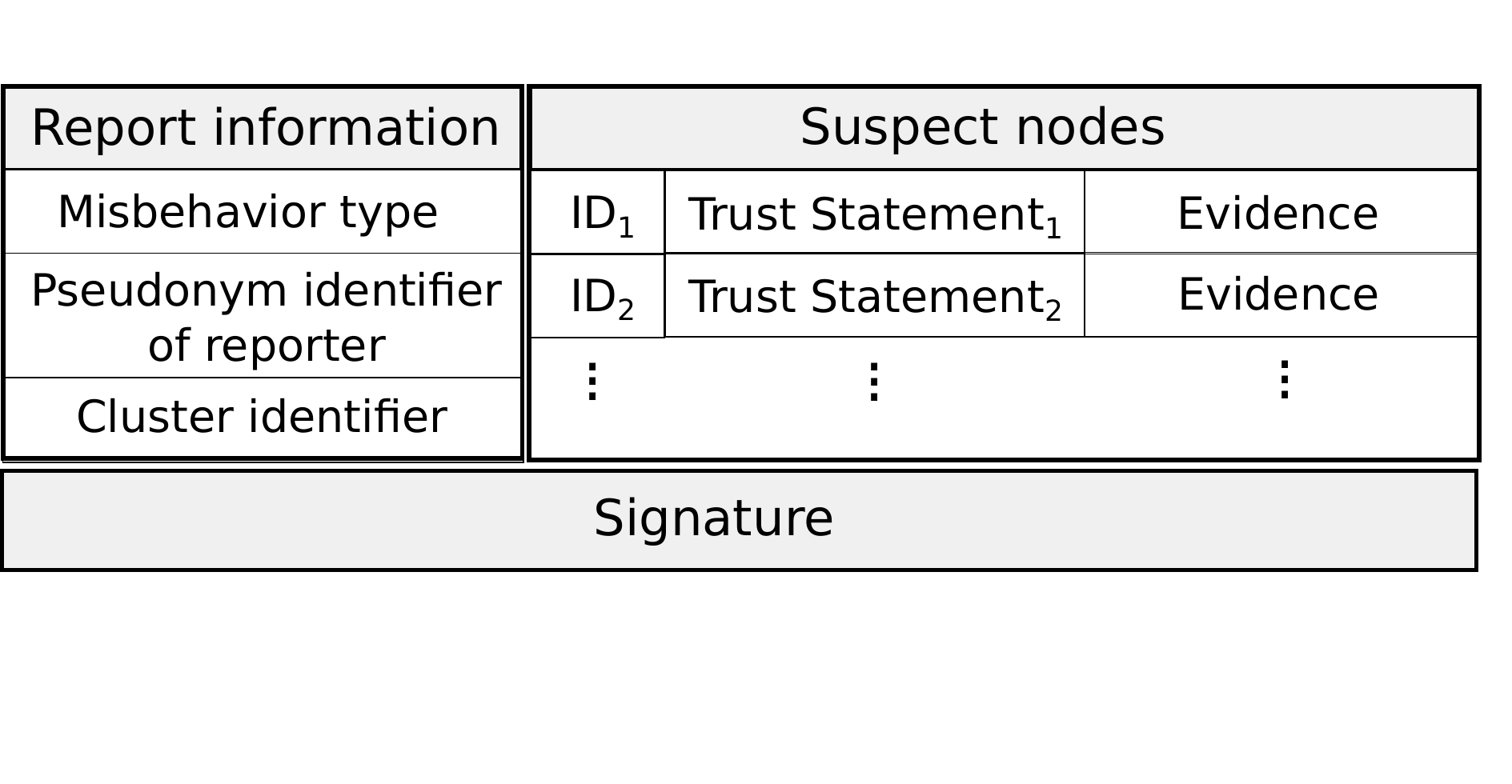}
	\caption{Sample structure of a misbehaviour Report}\label{fig:report}
\end{figure*}

One of the proposed standards to organize such a VPKI, proposed by Whyte~et~al.~\cite{Whyte-SCMS}, is the security credential management system (SCMS), which proposes a number of authorities to protect the privacy of the users.
Issuing pseudonyms involves the following authorities: the enrollment certificate authority (ECA), responsible for long-term identities of vehicles, the registration authority (RA), who essentially checks whether a vehicle may still receive pseudonyms, and the pseudonym certificate authority (PCA), which issues pseudonyms.
When vehicles or cluster heads report misbehavior, this report mainly concerns specific vehicles and includes evidence (see e.g.~\cite{Bissmeyer-Central}).
This evidence consists of signed messages, and potentially misbehavior detection results that can be verified by re-executing the detection process (i.e., they are publicly verifiable); a sample misbehavior report can be found in Figure \ref{fig:report}.
In this example, the trust statement mirrors the result of the local misbehavior detection system; apart from identifiers of the suspected vehicles, the report stores the detected misbehavior, the pseudonym identifier of the reporter and the associated cluster identifier. 
This information is processed by a misbehavior authority (MA), which can decide on the validity of these reports and subsequently revoke pseudonyms and long term identities in cooperation with the PCA and one or more linkage authority (LA).
This protocol also informs the RA not to issue any more certificates for the reported vehicle.

It is important that the MA only revokes based on reliable reports, which requires that the MA is able to validate the reports of vehicles (i.e., objectively check the evidence) and detect when an attack against the revocation system itself is on-going.
Attacks on the revocation system include those that exist for traditional reputation systems (e.g., bad-mouthing attacks, where an attacker creates false accusations), often combined with Sybil attacks (where an attacker uses multiple pseudonyms to artificially increase the evidence for their claim).
In our proposed system model, we allow the attacker to use at most two pseudonyms at any time, in order to limit the Sybil attack capabilities within a cluster.
This can be achieved in real-world system by limiting the validity of certificates appropriately (as discussed in EU proposals~\cite{Ma-Pseudonym}), increasing the overhead, but providing tighter control with limited privacy loss.

\begin{figure*}
    \centering
    \includegraphics[width=2\columnwidth]{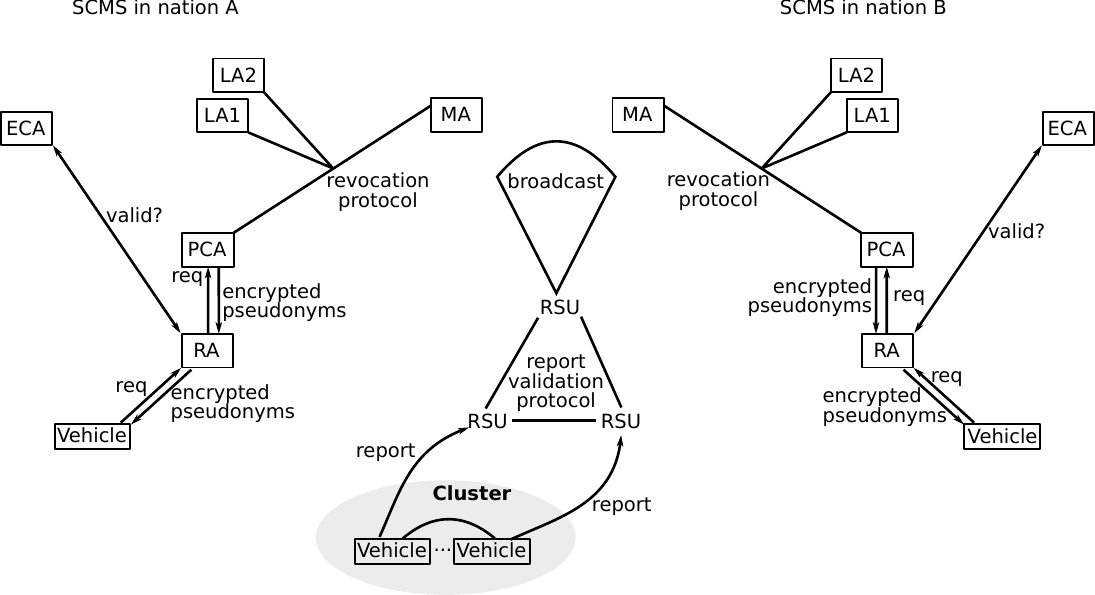}
    \caption{Blackchain's underlying system architecture.}\label{fig:arch}
\end{figure*}

\section{Blackchain}
\label{sec:blackchain}
We propose Blackchain (\emph{Black}box Block\emph{chain}), with which we aim to provide cluster-based VANETs with an integrated accountability system that exploits clusters to create a distributed ledger for exchanged messages.
Since these messages relate to real-world observations and processes, there are objective ways to establish which of these messages are correct (i.e., corresponding to the real world), and which contain false data.
Detecting malicious actors this way is referred to as \emph{misbehavior detection}: for a survey of different mechanisms that can be used for this purpose, we refer interested readers to our recent survey on this topic~\cite{vanderHeijden-Survey}.
This objective truth is publicly verifiable, i.e., it can also be used to detect attackers at a central location, such as the MA discussed in the previous section.
In this paper, we propose that the Blackchain can be used to perform this centralized misbehavior detection and revocation without requiring trust in any individual trusted third party (TTP).
The concept is shown in Figure \ref{fig:arch}: different countries will likely run their own SCMS, and a protocol is needed to perform cross-border revocation.
Our proposal not only enables this functionality, but also makes each SCMS accountable towards the participating vehicles, because Blackchain's publicly verifiable permissionless Blockchain can be accessed (and potentially mined) by anyone.

Each vehicle accumulates information about it's own state and, through received messages, about other vehicles in the vicinity.
Unlike the classical approach to store these state changes in an EDR, with the overhead of a trusted platform to ensure the append-only property, we persist these changes in a DL.
A direct approach to this would be to require each vehicle to participate in the Blackchain directly as a network node.
Having observations from different nearby neighbours in the Blackchain, malicious behaviour can easily be detected through misbehavior detection.
By propagating the resulting blocks to the MAs, who also participate in the Blackchain network, a consensus decision can then be made to revoke the corresponding vehicle, which can be stored in the Blackchain along with the associated evidence, persisting all the relevant information automatically.
This would result in a public, permissionless Blockchain, where all vehicles and MAs mine blocks by reaching consensus about misbehavior detection decisions.

However, this approach is not viable in the presence of millions of vehicles, the lack of permanent internet connectivity, and high message frequencies (10 Hz per vehicle).
Therefore, we reduce the amount of state updates, and remove the requirement that vehicles mine into the permissionless blockchain, by replacing it with a localized, permissioned chain.
To reduce the amount of state updates, we require vehicles to participate in clustered communication to agree on a common state, and to decide on potential revocation decisions (using a local revocation protocol, such as OREN~\cite{Bilogrevic-OREN}).
The cluster state reduces the size and frequency of updates from vehicles, but still allows other parties to verify the correct behaviour of the cluster participants.
These clusters are implemented using a permissioned Blockchain (i.e., only non-revoked vehicles and RSUs may participate), and their decisions are forwarded to the MAs.
Hierarchical clusters can be used to further reduce the size of these consensus groups, if necessary for performance reasons.

The advantage of using hierarchical clusters is that these will consist of RSUs, all of which have known identities (because such devices will not have privacy requirements anyway); thus, these can be used to run a byzantine fault tolerance (BFT) consensus.
This is not so easy within a cluster of vehicles and RSUs, because vehicles can potentially change their identity once per time slot (due to the partially overlapping validity periods discussed in the previous section). 
However, if all RSUs around the world are used in BFT consensus, this will still result in poor performance, due to latency and the sheer number of RSUs that some proposals suggest (some even recommend full road coverage).
To resolve this issue, we propose that BFT consensus may be reached in smaller groups of RSUs; unlike within vehicle clusters, latency requirements are not as significant an issue.
Clustering can be done by different parameters, depending on how widely RSUs are deployed, e.g., all RSUs in one city or all RSUs in a specific grid area.
These RSUs then report the resulting consensus and corresponding evidence to the MAs, which use these decisions as input for the public, permissionless Blockchain.
Since the consensus relies on the evidence provided by vehicles, the reports received by an MA can still be publicly verified; thus the public nature of the Blackchain allows all participants to verify the correctness of these decisions.
The transaction format on the public chain therefore contains the report with evidence for internal consistence, a signature, e.g.,  of the RSU group, and a list of references, identifying the participants with there identities in the public chain. This structure allows for a full audit up to the introduction of the participants, which signed the report, in the public chain. This introduction is again a consensus of the participants of the public chain.

\section{Conclusion}
\label{sec:conclusion}
In this paper, we have provided some conceptual foundations for Blackchain, a distributed ledger that provides accountability for misbehavior authorities and vehicles alike.
The purpose of Blackchain is to reduce the trust requirements on users of a vehicular communication system, improving the performance of global revocation algorithms by employing hierarchical consensus, and creating accountability for misbehavior authorities.
However, these foundations are only the first step in this area of research: there are many open questions that still need to be solved to make this system practically feasible.
From the vehicular perspective, the most important factor is whether clusters are stable enough to provide the necessary consensus algorithms.
From a distributed ledger perspective, the most exciting question is what guarantees hierarchical consensus can provide compared to a full consensus where all RSUs and MAs (and even potentially all vehicles) participate.
Although Blackchain itself may not be feasible to implement, we think our proposal gives interesting directions of research for both fields, which are valuable beyond the actual implementation of Blackchain.

\begin{acks}
This work was partially funded by the Baden-W\"urttemberg Stiftung. In addition, we would like to thank Benjamin Erb for the initial discussion that lead to the creation of this paper, and the anonymous reviewers for their insightful and helpful comments.
\end{acks}

\balance

\bibliographystyle{ACM-Reference-Format}
\bibliography{references}

\end{document}